# A multi-interval MBSC theory for active correlations technique

*Yu.S.Tsyganov*

Joint Institute for Nuclear Research, Dubna

**Abstract**

The purpose of the paper is the development of the formalism for the treatment of rare events especially, when one applies "active correlation" method to suppress background products in the heavy-ion induced complete fusion nuclear reactions. This formalism in fact is an extension of classical BSC formalism for the case of time multi intervals.

## 1. Introduction

In interpretation of rare detected events in the heavy-ion induced complete fusion nuclear reactions one should consider two basic approaches to the problem. One of them is formalized in the concept of a linked decay signal combination (LDSC) [1], which tests whether the analyzed signal sequence does fit this concept or not. Another is formalized in the concept of background signal combination (BSC) [2] and tests whether the analyzed signal sequence does fit this concept. In case of poor experimental statistics we do not know the structure of the decay chain a priori; neither the reliable information about half-lives of members of this chain is available. In this situation the second approach is inevitable and most natural, i.e., to look if the analyzed signal group corresponds to the BSC pattern or not and thereby get the answer to the basic question mentioned above.

Recently, a dramatic success was achieved in synthesis of SHE[1] [3]. It was the "active correlation technique" which allowed to detect rare decays with the radical suppression of background products [4-5]. The significant results were obtained using the Dubna Gas-Filled Recoil Separator (DGFRS). It is this facility that has allowed to synthesize a lot of new SHE isotopes with Z=110 to 118 [6]. The DGFRS detection system plays an important role in these experiments [7, 8].

## 2. Experimental technique

In the DGFRS experiments the $^{48}$Ca-ion beam was accelerated by the U400 cyclotron at the Flerov Laboratory of Nuclear Reactions. The ERs recoiling from the target were separated in flight from $^{48}$Ca ions by the DGFRS. ERs passed through a time-of-flight (TOF) system and were implanted in a 4x12 cm$^2$ semiconductor detector array with 12 vertical position-sensitive strips, located at the separator's focal plane (Fig.1). This detector was surrounded by eight 4x4 cm$^2$ side detectors without position sensitivity, forming a box open to the front (beam) side. The position-averaged detection efficiency for alpha decays of implanted nuclei was 87% of 4π. The detection system was tested by registering the recoil nuclei and decays (α and SF) of the known isotopes of No and Th, as well as their descendants, produced in the reactions $^{206}$Pb($^{48}$Ca,xn), and $^{nat}$Yt($^{48}$Ca,xn), respectively. The energy resolution for α-particles absorbed in the focal plane detector was ~ 50-95 KeV. The α-particles that escaped the focal-plane detector at different angles and registered in a side detector had an energy resolution of the summed signals of 140-220 KeV. The full width at half maximum (FWHM) position resolution of the signals of correlated decays of nuclei implanted in the detectors were 0.9-1.4 mm for ER-α signals and 0.5 – 0.9 mm for ER-SF signals. The detecting module of the DGFRS is shown in the Fig.1.

---

[1] Super Heavy Elements

The "VETO" detector placed behind the focal –plane one ion order to suppress long-path charged particles backgrounds.

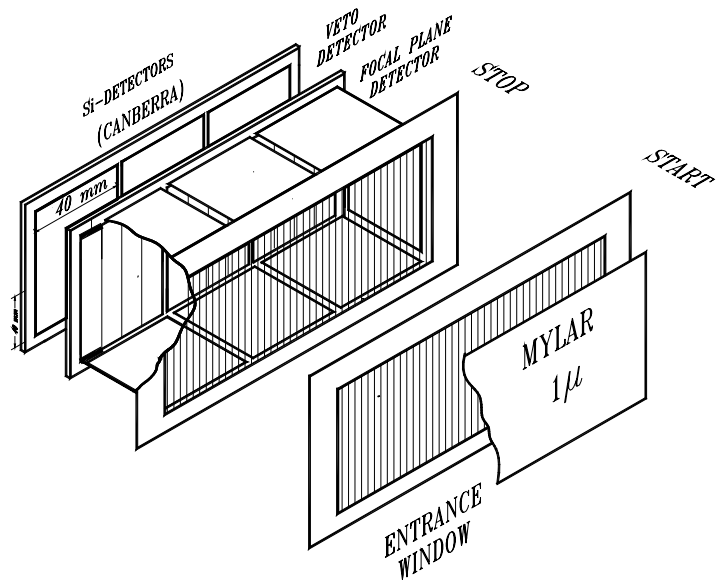

**Fig.1** Detecting module of the DGFRS. Focal plane, side, VETO and TOF (START and STOP) detectors are shown.

## 3. Real-time ER-α sequences detection mode

When applying very intense beams, one should take into account not only charged particle background products. Neutron induced background can not be suppressed by using TOF systems. Contributions of all backgrounds to the measured spectra, related with cyclotron, are negligible if one applies beam-off detection mode. Since the object under investigation usually yields genetically linked chains of more than one α-particle, one can apply the first recoil-alpha link as a pointer to a probable forthcoming event. The basic idea of the algorithm to perform such operation is aimed at searching in a real-time mode of time-energy-position recoil-alpha link using the discrete representation of the resistive layer of each position sensitive strip.

So, after beam is switching off, one has an opportunity to detect the forthcoming α-particle signals as well as SF-signals with a negligible in a background signals.

## 4 MBSC - multi interval approach

The basic parameter of BSC theory is the time duration between starter (e.g. recoil signal) and finisher (e.g. SF signal). Let us consider the composition of the "event" when a few strongly non random factors act within this time, like it is shown in the Fig. 2.

In this connection the whole "event" should be considered as a combination of random events of definite nature for all intervals, and the total probability value will be $P_{tot} = \prod P_{tot}^{i}(t_i, t_{i+1})$.

Here i – interval number. Of course, each $P_i$ value has it's own configuration probability value. Note, that for the case of practical use of electromagnetic separators, random event imitators of α-decays have a quite different nature.

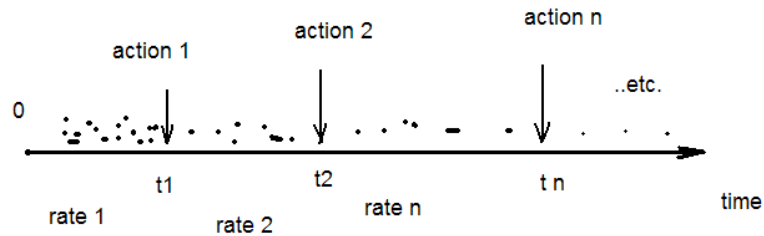

**Fig.2** A flowchart of a multi interval process with non random actions 1-n.
(rate $_{i+1}$ << rate $_i$)

## 5. Analysis of rare events in heavy-ion experiments: two intervals

Stochastic Poisson time processes which proved perfectly in many scientific application fields as models of random event streams while the necessary a priori information about their origin is absent, seem to be the best means for the description of background decays. These processes are the time functions $k(t_1,t_2)$ – numbers of random events, occurring during a time interval $t = t_2 - t_1$ with a probability $Q_k(t)$. The function of probability distribution of k(t) is given by the formula

from Ref.[2] as : $Q_k(t) = \dfrac{(lt)^k}{k!} \exp(-lt)$, $t \in (0, \infty)$,  (1)

where l is parameter of the Poisson distribution, t is time. The quantity of $lt$ is the expectation and at the same time the variance of k (t) at a moment t. In the case of the signals of different type the formula (1) describes only probabilities of their sums irrespectively of their order and configuration. If the event is made up of time-independent signal combinations of m-kinds, the probability distribution of the definite configuration is

$$Q_{sk}(t) = p_S \cdot \prod_{i=1}^{m} Q_{k_i}(t) , \qquad t \in (0, t_i ), \tag{2}$$

where $P_s$ is the probability of combination of s - th kind [2].

Considering the application the above mentioned technique, one can point out the following signals/imitators under investigation:

1 – ER, recoil of superheavy element implanted into the focal-plane detector and detected before by TOF detector;

2 - $\alpha$ - particle signal in the range of 8 -12 MeV, having normal y-position signal;

3 - the same as 2, but without position signal and detected by side detector;

4 - the same as 2, but detected in "beam OFF" pause;

5 - the same as 3, but in "beam OFF" pause;

6 - SF imitator signal.

In addition, according to [2] any combinations within the same group are considered to have no statistical significance.

So, the typical common combinations are:

1-222…2 – 444..4 -55..5 -6 , etc.

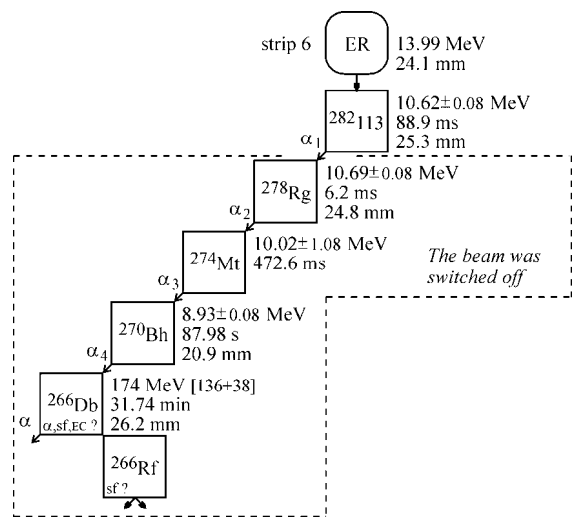

**Fig.3** Z=113 multi-chain event measured in [9].

Let us consider the two-interval combination when N – the total number of signals in the whole chain (see Fig.2), n3 – number of signals of type 3, n4 – type 4, n5 – type 5 ( n2+n3+n4+n5 =N-2).

Analogous to [2] it is easy to obtain a configuration factor $p_s$ in the form of

$P_{tot} = P_{t1} \times P_{t2}$, with two configuration probabilities $P_{s1}$ and $P_{s2,}$ to which correspond to beam–on and beam-off intervals.

$$P_{S_1} = \frac{n_2! n_3!}{(1+n_2+n_3)!}, \quad \text{and} \quad P_{S_2} = \frac{n_4! n_5!}{(n_4+n_5+1)!} \qquad (3)$$

$$P_{total} = P_{S1} \cdot P_{S2} \cdot \prod_{i=1}^{3} \frac{(l_i t_1)^{n_i}}{n_i!} \exp(-l_i t_1) \cdot \prod_{i=4}^{6} \frac{(l_i t_2)^{n_i}}{n_i!} \exp(-l_i t_2) \qquad (4).$$

Here: $n_1$=1 (starter), and $n_6$ =1 (finisher). Moreover, for typical DGFRS experiments $n_3$ is often equal to zero, when the first recoil-alpha sequence stops the cyclotron beam for a definite time.

Let us consider an example for real case of detecting chain of Z=113 decay [9]. Measured parameters of decay are: N=6, $n_1$ =1, $n_2$ =1, $n_3$ =0, n4 = 2, n5 =1, $n_6$=1.

Rates per $3\sigma^2$ of vertical position resolution element are:

$l_1$ (recoils) = 0.042 s$^{-1}$; $l_2$ ($\alpha$ in beam) = 0.0266; $l_4$ ($\alpha$ out of beam) =0.000025; $l_5$ (no position) =0.0018; $l_6$ (SF, beam off) = 2·10$^{-6}$.

In this connection: $P_{S1}$=0.5; $P_{S_2} = \frac{2! \cdot 1!}{(2+1+1)!} = \frac{2}{4!} = \frac{1}{12} = 0.0833$.

Considering two beam ON/OFF time intervals one can easily calculate the final probability value, taking into account two configuration values calculated above. So[3],

$$P_{tot} = 0.5 \cdot 0.0833 \cdot \frac{l_R t_{in}}{1!} \cdot \exp(-l_R t_{in}) \cdot \frac{(l_\alpha t_{OFF_\alpha})^2}{2!} \cdot \exp(-l_\alpha t_{OFF_\alpha}) \cdot \frac{l_{SF} t_{OFF_{SF}}}{1!} \cdot \exp(-l_{SF} t_{OFF_{SF}} \cdot \ln 2) =$$

6.2·10$^{-15}$, and corresponding number of random chains should be considered approximately as 9.3·10$^{-13}$.

Note, that the last value is in a good agreement with the same parameter that has been reported in [9].

---

[2] +/- three standard deviations
[3] $t_{in}$ and $t_{OFF}$ are calculated according to [2]

## 6. Summary

Well known BSC theory for ultra rare event detecting and interpretation is extended to the form of MBSC approach, which corresponds to the "active correlations" technique applied for radical suppression of beam-associated background. Formulae for calculation of the probability values are proposed. A reasonable field of application of MBSC theory is synthesis of SHE in heavy-ion induced complete fusion nuclear reactions.

This work is supported in part by RFBR Grant №07-02-00029.

**Supplement1: regular beam modulation**

In the case of regular beam modulation of millisecond duration one can consider a detected multi event as background signals following to the native one interval consideration [2]. In this case a typical sequence will be in the form of: r – a -α- α - a-.. , where the symbol "α" denotes beam-off alpha decay, whereas symbol "a" denotes in- beam alpha decay imitator.
If some decay contains n alphas in beam and k alphas out of beam, one can easily apply BCS formula for the calculation.

Namely, $P_{tot} = \frac{n!k!}{(n+k+1)!} \cdot (l_{ER}t) \cdot \exp(-l_{ER}t) \cdot \frac{(l_a t)^n}{n!} \cdot \exp(-l_a t) \cdot \frac{(l_\alpha t)^k}{k!} \cdot \exp(-l_\alpha t)$,

$$P_{tot} = \frac{l_{ER}t \cdot \exp(-l_{ER}t) \cdot (l_a t)^n \cdot \exp(-l_a t) \cdot (l_\alpha t)^k \cdot \exp(-l_\alpha t)}{(n+k+1)!}.$$

In particular, if no beam-off phase, when k≡ 0,

$P_{tot}^{in} = \frac{l_{ER}t \cdot \exp(-l_{ER}t) \cdot (l_a t)^N \cdot \exp(-l_a t)}{(N+1)!}$, where N - number of α-decay imitators.

**Supplement 2: balance equation for a single starter to switch a beam off**

The declared equation follows from the equivalence of the probabilities for the cases of single and pair correlation starters.
That is:

$(l_R t_R) \cdot \exp(-l_R t_R) = \frac{1}{2!} \cdot (l_R t_{R-\alpha}) \cdot \exp(-l_R t_{R-\alpha}) \cdot (l_\alpha t_{R-\alpha}) \cdot \exp(-l_\alpha t_{R-\alpha})$, or

$t_R \cdot \exp(-l_R t_R) = 0.5 \cdot t_{R-\alpha} \cdot \exp(-l_R t_{R-\alpha}) \cdot (l_\alpha t_{R-\alpha}) \cdot \exp(-l_\alpha t_{R-\alpha})$